\documentclass[12pt]{article}
\usepackage{amssymb}
\usepackage{subfigure}
\usepackage{graphicx}
\usepackage{float}

\usepackage{amsmath}
\usepackage{color}

\newcommand{\bb}{\begin{equation}}
\newcommand{\ee}{\end{equation}}
\newcommand{\bega}{\begin{eqnarray}}
\newcommand{\ega}{\end{eqnarray}}
\newcommand{\begae}{\begin{eqnarray*}}
\newcommand{\egae}{\end{eqnarray*}}

\newcommand{\h}{\hspace*{4ex}}

\newcommand{\cent}{\centerline}
\newcommand{\vs}{\vspace*}

\begin{document}

\baselineskip 0.5cm

\begin{center}

{\large {\bf Experimental optical trapping of microparticles with an Airy beams array using a Holographic Optical Tweezer} }

\end{center}

\vs{0.2 cm}

\cent{Rafael A. B. Suarez$^{\: 1}$, Antonio A. R. Neves$^{\: 1}$, and Marcos R. R. Gesualdi$^{\: 1}$}

\vs{0.2 cm}

\centerline{{\em $^{\: 1}$ Universidade Federal do ABC, Av. dos Estados 5001, CEP 09210-580, Santo Andr\'e, SP, Brazil.}}

\vs{0.5 cm}

{\bf Abstract  \ --} \ In this work, we present the experimental optical trap of microparticles with an Airy beams array using a holographic optical tweezers. The Airy beams array are attractive for optical manipulation of particles owing to their non--diffracting and autofocusing properties. An Airy beams array is composed of $N$ Airy beams which accelerate mutually and symmetrically in opposite direction, for different ballistic trajectories, that is, with different initial launch angles. Based on this, we developed a holographic optical tweezers system for the generation of non-diffracting beams and with it, we investigate the distribution of optical forces acting on microparticles of an Airy beams array. The results show that the gradient and scattering force of array on microparticles can be controlled through a launch angle parameter of Airy beams. In addition, it's possible to obtain greater stability for optical trap using an Airy beams array, with interesting possibilities for trapping and guiding of microparticles in a controllable way that can be applied in optical, biological and atmospheric sciences. \\


\vs{0.5 cm}

\h {\em\bf 1. Introduction} 

In 1986  A. Ashkin \textit{et al}. were able to capture in three--dimensional dielectric particles using a single--beam tightly focused by a high numerical aperture lens. This technique is now referred to as ``optical tweezers'' or ``optical trapping''~\cite{Askin1986,Askin2000}. The measurement of forces on micro--sized particles of picoNewton order with high precision, optical tweezers has become a powerful tool for application in different fields of research, mainly in the manipulation of biological system~\cite{Wang1997,Chiou2005}, colloidal systems~\cite{yada2004direct}, in nanotechnology for trapping of nano--structures~\cite{Marago2003}, in optical guiding and trapping of atoms \cite{Kuga1997} as well as the study of mechanical properties of polymers and biopolymers~\cite{sun2001stretching}. \\

On the other hand, the study of non--diffracting waves or diffraction--resistant waves in optics are special optical beams that keep their intensity spatial shape during propagation. Non--diffracting beams include Bessel beams, Airy beams and others~\cite{durnin1988comparison,Sivilo2007,Siviloglou2007,Suarez2016}; as well as the superposition of these waves can produce very special structured light beams~\cite{Vieira2012,Pachon2016,suarez2019}. These special optical beams present very interesting properties and could be applied in many fields in optics and photonics.
\\

Particularly, the Airy beams (AiBs) has attracted great interest recently in optical tweezers, for trapping and guiding of micro and nano--particles~\cite{Baumgartl2008,Cheng2010,Yang2012,Zhao2016,Lu2017a,Suarez2020}, due to their unusual features such as the ability to remain diffraction-free over long distances while they tend to freely accelerate during propagation~\cite{Sivilo2007,Siviloglou2007,Suarez2016}. An important parameter in the dynamic propagation is the initial launch angle which can be used to obtain optimal control of the ballistic trajectory~\cite{Siviloglou2008,Hu2010,suarez2020propagation}. Recently, several authors have addressed the propagation properties of AiBs to study of circular Airy beams (CAiBs) \cite{Efremidis2010,Papazoglou2011,Jiang2018} and radial array Airy beams~\cite{Vaveliuk2014,Chen2014} for optical trapping due to its unique abruptly auto--focusing characteristics where optical spot obtained in the focal field can be used for simultaneous trapping of multiple particles \cite{Zhang2011,Jiang2013,Jiang2016,Zhang2013,Cheng2014}.
\\

The introduction of holographic optical elements in the optical tweezers configuration has certainly become an advantage in the process of trapping, moving and manipulating microparticles with high precision~\cite{curtis2002dynamic,brzobohaty2013optical,Sun2008}. The use of computer--generated holograms and diffractive optical elements, such as spatial light modulators, allowed the simultaneous creation of several possibilities  of the optical traps, each with its unique characteristics~\cite{dufresne2001computer,bhebhe2018vector}. Such as trapping systems are called holographic optical tweezers. In holographic optical tweezers, the wavefront control allows for easy three-dimensional positioning of traps, as well as the creation of special optical, beams such as the Bessel, Laguerre-Gauss, Airy beams and a specific superposition of these beams, generating new special structured optical beams~\cite{Zhang2011,Zhang2013,bhebhe2018vector,suarezexperimental2020}.
\\

In this work, we present an experimental optical trap of microparticles with an Airy beams array using a holographic optical tweezers. An Airy beams array is composed of $N$ Airy beams which accelerate mutually and symmetrically in opposite direction, for different ballistic trajectories, that is, with different initial launch angles. Based on this, we investigate the optical force distribution acting on microparticles using an optimized experimental holographic optical tweezers system. This system presents very interesting possibilities for static and dynamical applications in optical trapping, optical guidance and optical manipulation.
\\

\h {\em\bf 2. Non-diffracting Airy beams} 

\textbf{Airy beams (AiBs):} The solution for AiBs propagating with finite energy can be obtained by solving the normalized paraxial equation of diffraction in $1$D~\cite{Sivilo2007} 
\begin{equation}
i\frac{\partial}{\partial \xi}\psi\left(s,\xi \right)+\frac{1}{2}\frac{\partial^{2}}{\partial s^{2}}\psi\left(s,\xi \right)=0\,,
\label{paraxial_equation}
\end{equation}
where $\psi$ is the scalar complex amplitude, $s=x/x_{0}$ and  $\xi=z/kx_{0}^{2}$ are the dimensionless transverse and longitudinal coordinates, $x_{0}$ its characteristic length and $k=2\pi n/\lambda_{0}$ is the wave--number of an optical wave. The Eq.~(\ref{paraxial_equation}) admits a solution at $\xi=0$, given by~\cite{Siviloglou2008}
\begin{equation}
\psi\left(s,0 \right)=\text{Ai}\left(s \right)\text{exp}\left(as \right)\text{exp}\left(i\nu s \right)\,,
\label{Airy_finity}
\end{equation} 
where $\text{Ai}$ is the Airy function, $a$ is a positive quantity which ensures the convergence of Eq.~(\ref{Airy_finity}), thus limiting the infinity energy of the AiBs and $\nu$ is associated with the initial launch angle of this beam. The scalar field $\psi\left(s,\xi \right)$ is obtained from the Huygens--Fresnel integral, which is highly equivalent to Eq.~(\ref{paraxial_equation}) and determines the field at a distance $\xi$ as a function of the field at $\xi=0$~\cite{Siviloglou2008}, that is 
\begin{equation}
\begin{split}
\psi \left( s, \xi \right)=\text{Ai}&\left(s-\dfrac{\xi^{2}}{4}-\nu \xi + ia\xi\right)\text{exp}\left[ a \left(s - \dfrac{\xi^2}{2} - \nu \xi \right)\right] \\
& \times  \text{exp}\left[ i \left( - \dfrac{\xi^3}{12} + \left(a^2 -\nu^2 + s \right)\dfrac{\xi}{2}  + \nu s -\nu \dfrac{\xi^2}{2} \right)\right].
\end{split}
\label{Airy_1D}
\end{equation}

This equation shows that the intensity profile decays exponentially as a result of modulating it with a spatial exponential function on the initial plane $\xi=0$. The term $s_{0}=s-\left( \xi^{2}/4\right)-\nu \xi$, where $s_{0}$ denotes the initial position of the peak at $\xi=0$, defines the transverse acceleration of the peak intensity of AiBs.
\\

These results can be generalized for $2$D taking the scalar field of a beam described as the product of two independent components~\cite{Sivilo2007,Siviloglou2008}, that is:
\begin{equation}
\psi\left(s_{x},s_{y},\xi_{x},\xi_{y}\right)=\psi_{x}\left(s_{x},\xi_{x}\right)\psi_{y}\left(s_{x},\xi_{y}\right)\,,
\label{Airy_2D}
\end{equation}
where each of the components $\psi_{x}\left(s_{x},\xi_{x}\right)$ e $\psi_{y}\left(s_{x},\xi_{y}\right)$ satisfies the Eq.~(\ref{paraxial_equation}) and is given by Eq.~(\ref{Airy_1D}), with $s_{x}=x/x_{0}$, $s_{y}=y/y_{0}$, $\xi_{x}=z/kx_{0}^{2}$ e $\xi_{y}=z/ky_{0}^{2}$.
\\

\textbf{Airy beams array (AiBA):} Through a rotation of $\theta=2\pi/N$ of Eq.~(\ref{Airy_2D}) in the transverse plane, we obtain a set of $N$ rotated AiBs. These evenly displaced AiBs on the transverse plane in $(\delta x,\delta y)$, accelerate mutually in the opposite direction~\cite{Suarez2020,Lu2017}. The AiBA can be described as
\begin{equation}
\Psi(s_{x},s_{y},\xi_{x},\xi_{y})=\sum_{j=1}^{N}\psi_{jx}\left(s_{jx},\xi_{x}\right)\psi_{jy}\left(s_{jy},\xi_{y}\right) \,, 
\label{matriz_superpositions}
\end{equation}
where the dimensionless transverse coordinates are given by
\begin{equation}
s_{jx}=\frac{(x\cos\theta_{j}-y\sin\theta_{j}+\delta x)}{w_{0}}\,, \quad s_{jy}=\frac{(x\sin\theta_{j}+y\cos\theta_{j}+\delta y)}{w_{0}}\,,
\label{matriz_superpositions1}
\end{equation}
The angle $\theta_{j}=2(j-1)\pi/N$ denotes the angle of rotation around the $z$ axis. To simplify the AiBA description, we will consider a symmetrical configuration, such as, $a_{x}=a_{y}=a$, $\nu_{x}=\nu_{y}=\nu$ and $x_{0}=y_{0}=w_{0}$, resulting in $\xi_{x}=\xi_{y}=\xi=z/kw_{0}^{2}$.

\begin{figure}[H]
\centering
 \includegraphics[scale=0.60]{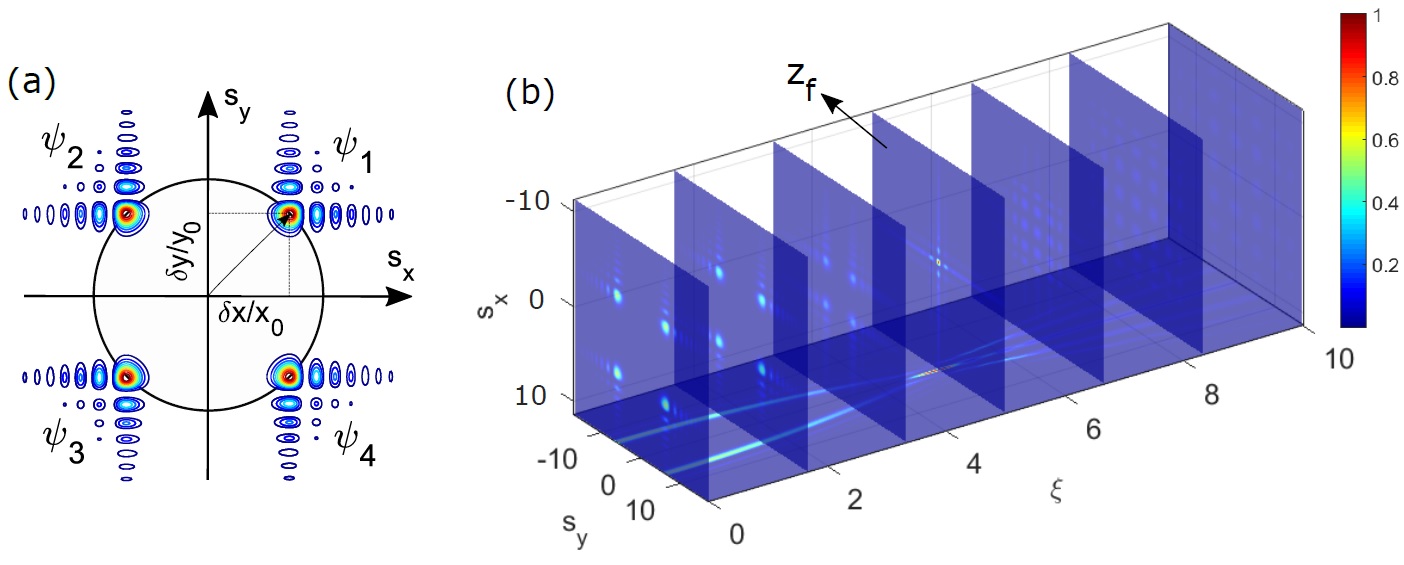}
\caption{$(a)$ Scheme for the generation of an AiBAs with four AiBs ($N=4$). $(b)$ Dynamic propagation of the normalized intensity distribution. The plane $z_{f}$ correspond to the focal plane.}
\label{Rotation}
\end{figure}

The focal point $z_{f}=2kw_{0}(\sqrt{\nu^2+x_{p}/w_{0}} - \nu )$, is defined by the common point where the main lobes from all the beams in the array intercept along of the beam, where $x_{p}$ is the position of the first intensity peak. In the Fig.~\ref{Rotation} we can see a representative scheme of the generation process from four AiBs propagating along the $z$--axis. \\

\textbf{Propagation of an AiB through ABCD paraxial optical system}. 
The propagation of an AiB one--dimensional through an ABCD paraxial optical system can be described by the generalized Huygens--Fresnel diffraction integral, which is given by~\cite{Morris2007}
\begin{equation}
\psi\left(x,z \right)=\dfrac{1}{\sqrt{i\lambda B}}\int_{-\infty}^{\infty} \psi(\eta,z=0)\text{exp}\left[\dfrac{ik}{2B}\left(Dx^{2}-2\eta x +A\eta^{2} \right) \right] d\eta\,,
\label{generalized_Huygens_Fresnel_diffraction_integral}
\end{equation}
where $\psi(x,z=0)$ and $\psi(\eta,z=0)$ are the fields in the input and output planes, respectively. After some calculations, we get an analytical expression for the AiB
\begin{equation}
\begin{split}
\psi\left(x,z \right)&=\dfrac{1}{\sqrt{A}}\text{exp}\left(\dfrac{ikC}{2}x^2 \right)\text{exp}\left[a\left(\dfrac{x}{Ax_{0}}-\dfrac{B^2}{2A^2k^2x_{0}^4}-\dfrac{\nu B}{Akx_{0}^2}\right)\right]\\
&\times \text{exp}\left\lbrace i\left[- \dfrac{B^3}{12A^3 k^3 x_{0}^6} + \left(a^2 - \nu^2 +\dfrac{x}{Ax_{0}} \right)\dfrac{B}{2Akx_{0}^2} + \dfrac{\nu x }{Ax_{0}}-\dfrac{\nu B^2}{2A^2k^2x_{0}^4}\right] \right\rbrace \\
&\times Ai \left(\dfrac{x}{Ax_{0}}-\dfrac{B^{2}}{4A^2k^{2}x_{0}^4}-\dfrac{\nu B}{Akx_{0}^2}+\dfrac{ia B}{Akx_{0}^2} \right).
\end{split}
\label{Airy_ABCD}
\end{equation}  

This result reveals that the Airy beams conserve your properties when it passed through an optical system but the characteristic parameters as dimensions of the transverse intensity pattern, the characteristic length and curvature (deflection) of the Airy beam can be modified~\cite{suarez2020propagation}.
\\

For a set of $N$ rotated AiBs, the Eqs.~(\ref{matriz_superpositions1}) became $s_{jx}=(x\cos\theta_{j}-y\sin\theta_{j}+ A\delta x)/w_{0}$ y $s_{jy}=(x\sin\theta_{j}+y\cos\theta_{j} + A\delta y)/w_{0}$~\cite{suarez2020propagation}.
\\

\h {\em\bf 3. Holographic optical tweezers} 

\textbf{Holographic computational method for Airy beam generation}
The holographic computational methods are now a well--established technique for generation and characterization of special optical beams and structured light, particularly non--diffracting beams~\cite{Vieira2012,Suarez2016,Pachon2016,Yepes2019}. The computer--generated holograms of these special beams are calculated and implemented in spatial light modulators and reconstructed optically in a holographic setup. These methods have generated experimental results of high quality and fidelity optical beams reconstruction compared to the theoretical predicted, because the holographic technique is an extremely accurate tool in the reconstruction of amplitude and phase of optical waves~\cite{Vieira2012,Suarez2016,suarez2019}. Particularly, in this work from the field that described the Airy beams Eq.~\ref{Airy_1D}, we build a Computer-Generated Hologram (CGH) which is optically reconstructed using a Spatial Light Modulator (SLM). The computer--generated hologram is calculated using an amplitude function which consists in varying the coefficient of transmission or reflection of the medium from the following amplitude function
\begin{equation}
H\left( x,y\right)= \frac{1}{2}\left\lbrace \beta\left( x,y\right) +a\left( x,y\right)cos\left[\phi\left( x,y\right)-2\pi \left(\xi x + \eta y \right)  \right] \right\rbrace \,,
\label{transmission}
\end{equation}
where $a\left( x,y\right)$ is the amplitude and $\phi\left( x,y\right)$ is a phase of the complex field, $\left(\xi,\eta \right)$ is a spatial frequency of the plane wave using as reference and $\beta\left( x,y\right)=\left[1+a^{2}\left(x,y \right)\right]/2 $ is the function bias taken as a soft envelope of the amplitude $a\left( x,y\right)$. The plane wave of reference is off-axis and introduces frequencies that separate the different orders of the encoded field~\cite{Vieira2012,Suarez2016,suarez2019}. \\

\begin{figure}[H]
 \centering
  \includegraphics[scale=0.70]{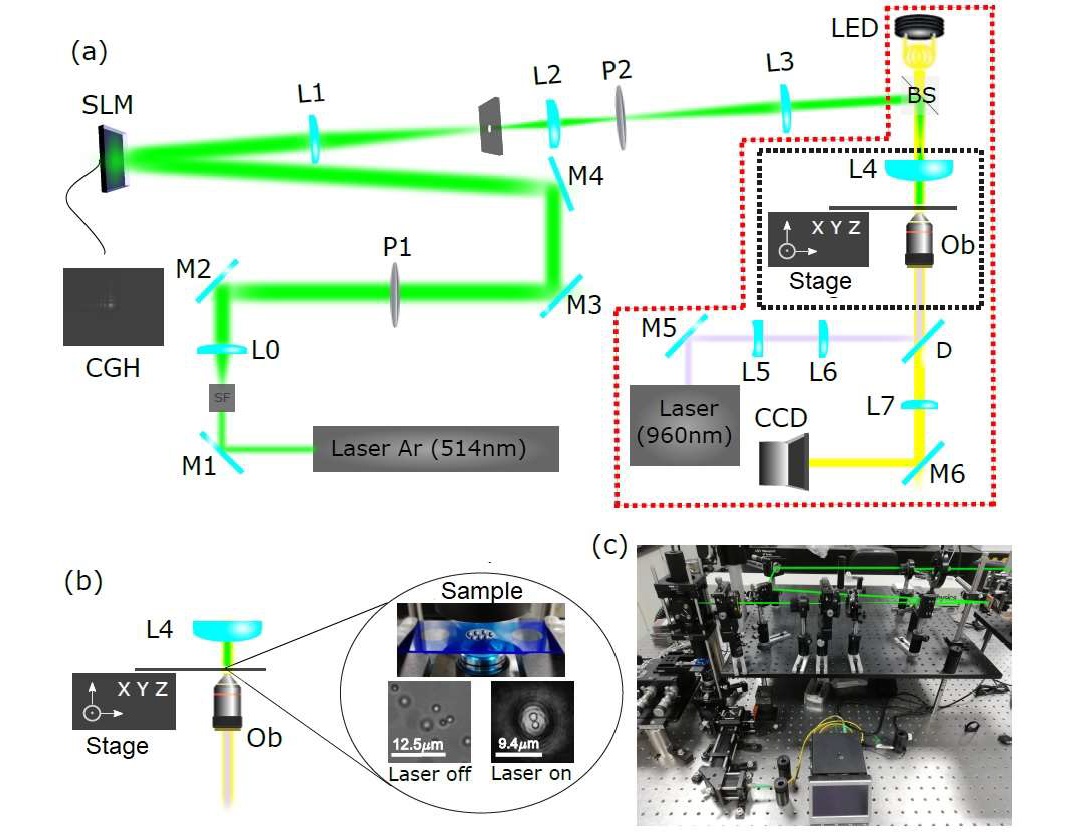}
 \caption{(a) Experimental setup for optical trapping using non--diffracting beams. SF is a spatial filter, Ls are lenses, Ms are mirrors, Bs is a beam splitter, Ps are polarizers, Ob is a high numerical aperture lens, D is the dichroic mirror, SLM is the reflection modulator and CCD is the camera for image acquisition; (b) Sample plate; and, (c) Holographic Optical Tweezers prototype}
 \label{Setup_Pinza}
\end{figure}

\textbf{Holographic optical tweezers setup}. For the development of the holographic optical tweezers prototype, shown in Figure~\ref{Setup_Pinza}, was used an Argon laser with wavelength $\lambda = 514.5 \text{nm}$ and output power of $ 300 \text{mW} $. Initially, the beam passes through the spatial filter where it is expanded and then collimated by the lens L1 with a focal length of $ 75 \text{mm} $ which results in a beam in diameter of about $ 10 \text{mm} $.  The beam is directed by the M2, M3 and M4 mirrors to the SLM (LETO, Holoeye Photonics), with each pixel measuring $6.4 \mu \text{m}$ in a display matrix $1920$ $\times$ $1080$.

In this system, to obtain an amplitude modulation of the Airy beams, a computer--generated hologram (CGH)  was employed, Eq.~\ref{transmission}, and it's implemented in the SLM. The optical axis of the polarizer P1 is aligned at an angle of $ 0^{0} $ and of the polarizer P2 at $90^{0} $ with respect to the $y$ axis of SLM. The hologram is reconstructed due to the beam diffraction in the hologram CGH in SLM. The reconstructed beam passes through two $4f$ systems in order to select the SLM diffracted beam and reduce the beam size respectively. 
\\

The first $4f$ system consists of two lenses, L1 and L2 of focal lengths $150\text{mm}$ and $50 \text{mm}$, respectively. In the focal plane of the L1 lens a mask was placed to select the different diffraction orders in the Fourier plane of the holographically reconstructed beam. The second $ 4f $ system was formed with two L3 and L4 focal length lenses $150\text{mm}$ and $25\text{mm}$, respectively. A beam splitter (BS) to reflect the beam vertically for the sample. The sample is placed in the focal plane of the L4 lens. To form the trapped particle image, the LED illumination system was used along with a $100\times $ and $ NA=1.25$ microscope objective.\\

\textbf {Optical tweezers calibration and Digital Video Microscopy}. With the experimental arrangement implemented, it was possible to capture and manipulate microparticles using the holographic optical tweezers setup with non-diffracting beams (in $ 514.5 \text{nm} $). To quantify the optical force, we need to monitor the position of the Brownian particles \cite{Jones2015, volpe2013simulation}. Knowing the trajectory of one of the trapped particles it is possible to calibrate the optical trap to obtain the value of the elastic constant ($k$).

\begin{figure}[H]
 \centering
  \includegraphics[scale=0.60]{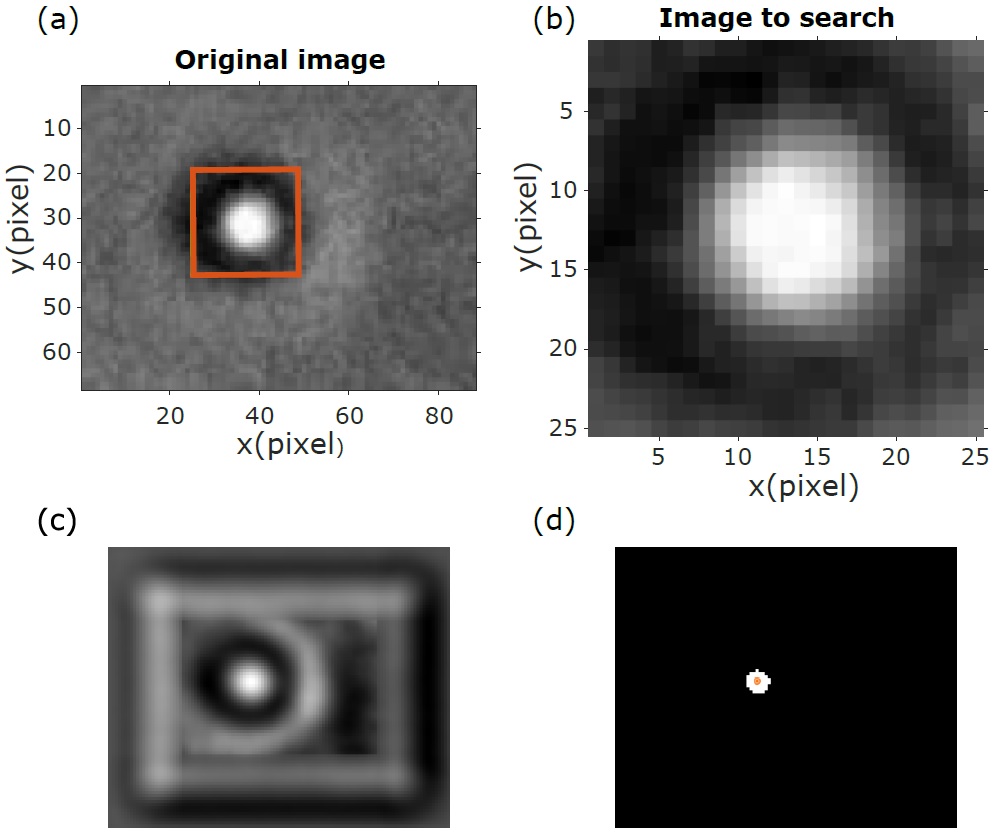}
 \caption{$ (a) $ Digital image $ (68 \text{pixel} \times 88 \text{pixel}) $ taken from a trapped polystyrene particle $ (a = 1.03 \mu m) $. $ (b) $ Model image $ (25 \text {pixel} \times 25 \text {pixel}) $, taken from original image, to calculate cross correlation. $ (c) $ Resulting array. $ (c) $ centroid corresponding to the resulting matrix.}
 \label{Microscopia_Digital}
\end{figure}

An important aspect of optical trapping and manipulation is the presence of Brownian motion where the microparticles that are suspended in a fluid undergo random movement due to collisions with the fluid molecules \cite{Jones2015, volpe2013simulation}. Therefore, the movement of an optically trapped particle is the result of random motion and the optical forces due to beam intensity gradients. 
\\

The process of measuring the Brownian motion of an optically trapped particle consists of recording a video and tracking its position frame by frame. This technique is commonly referred to as \textit{tracking}~\cite{helgadottir2019digital}, where each frame of the video is a two--dimensional digital image with a pixel matrix structure. For grayscale image, each pixel is associated with a value representing light intensity and can take values between $0$ (black) up to $255$ (white). 
\\

Figure \ref{Microscopia_Digital} $(a)$ shows the image obtained from a polystyrene particle of radius $ R = 1.03 \mu \text{m} $, trapped with a Gaussian laser beam of wavelength $ 975 \text{nm} $, Figure~\ref{Setup_Pinza}. Although there are many ways to execute tracking~\cite{Jones2015}, in this case we use a function that allows you to calculate normalized cross--correlation between a model matrix, Figure~\ref{Microscopia_Digital} $(b)$, and an original matrix, Figure~\ref{Microscopia_Digital} $(a) $, where the resulting matrix contains the correlation coefficients, Figure~\ref{Microscopia_Digital} $(c)$. The correlation is calculated between pairs of blocks belonging to the model image and the original image looking for the maximum in this measure. Finally, particle positions can be calculated as the centroids of this region, Figure~\ref{Microscopia_Digital}~$(c)$. This technique allows you to achieve sub--pixel resolution, up to one--tenth of the pixel size in the  $x$ and  $y$ directions~\cite{Jones2015}. 
\\
 
\begin{figure}[H]
 \centering
  \includegraphics[scale=0.60]{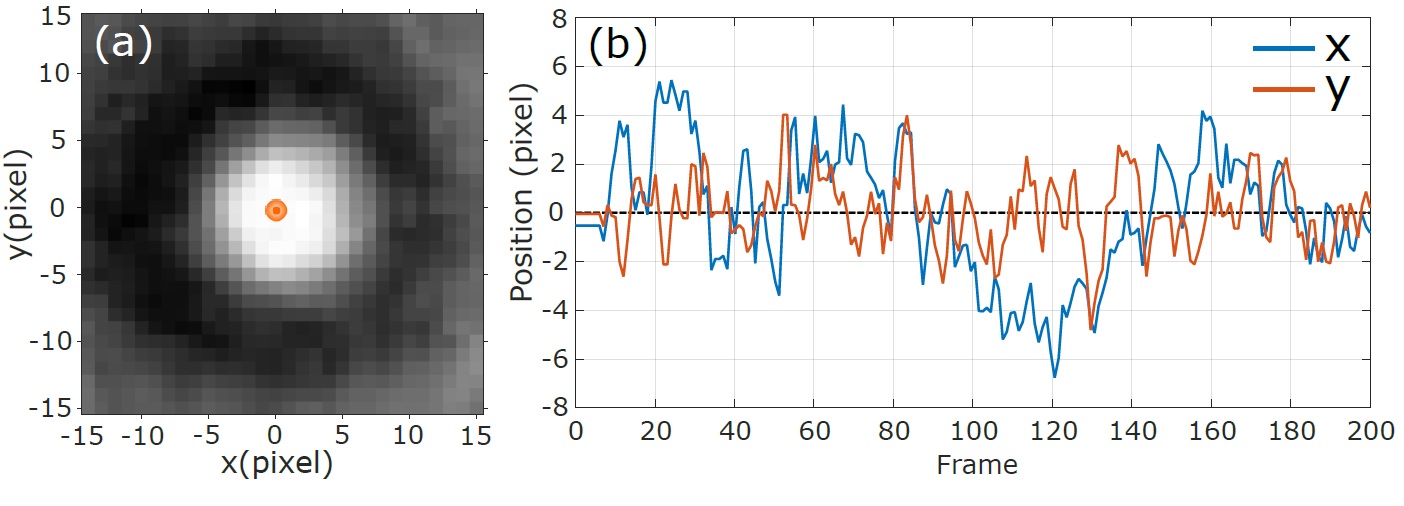}
 \caption{\textit{Tracking} of a trapped particle using digital video microscopy. $ (a) $ image of trapped polystyrene particle in equilibrium position. $ (d) $ Expected offset around the equilibrium position along the directions - $ x $ and - $ y $.}
 \label{Valor_medio_Microscopia}
\end{figure}

Once the centroids of each of the frames are calculated, it is necessary to join in sequence to define the trajectory. For this, it is necessary to establish as parameter the expected displacement between frame by frame of the particle. In Figure~\ref{Valor_medio_Microscopia} can be seen a microparticle trapped over the transverse plane. The corresponding expected offset (in pixel) along the - $ x $ and - $y$ directions as a function of the number of frames is shown in Fig.~\ref{Valor_medio_Microscopia} $(a)$. In this example, the video obtained was acquired at a rate of $185 \text{Hz}$ so the time between each frame is $5.4\text{ms}$.

\begin{figure}[H]
 \centering
  \includegraphics[scale=0.60]{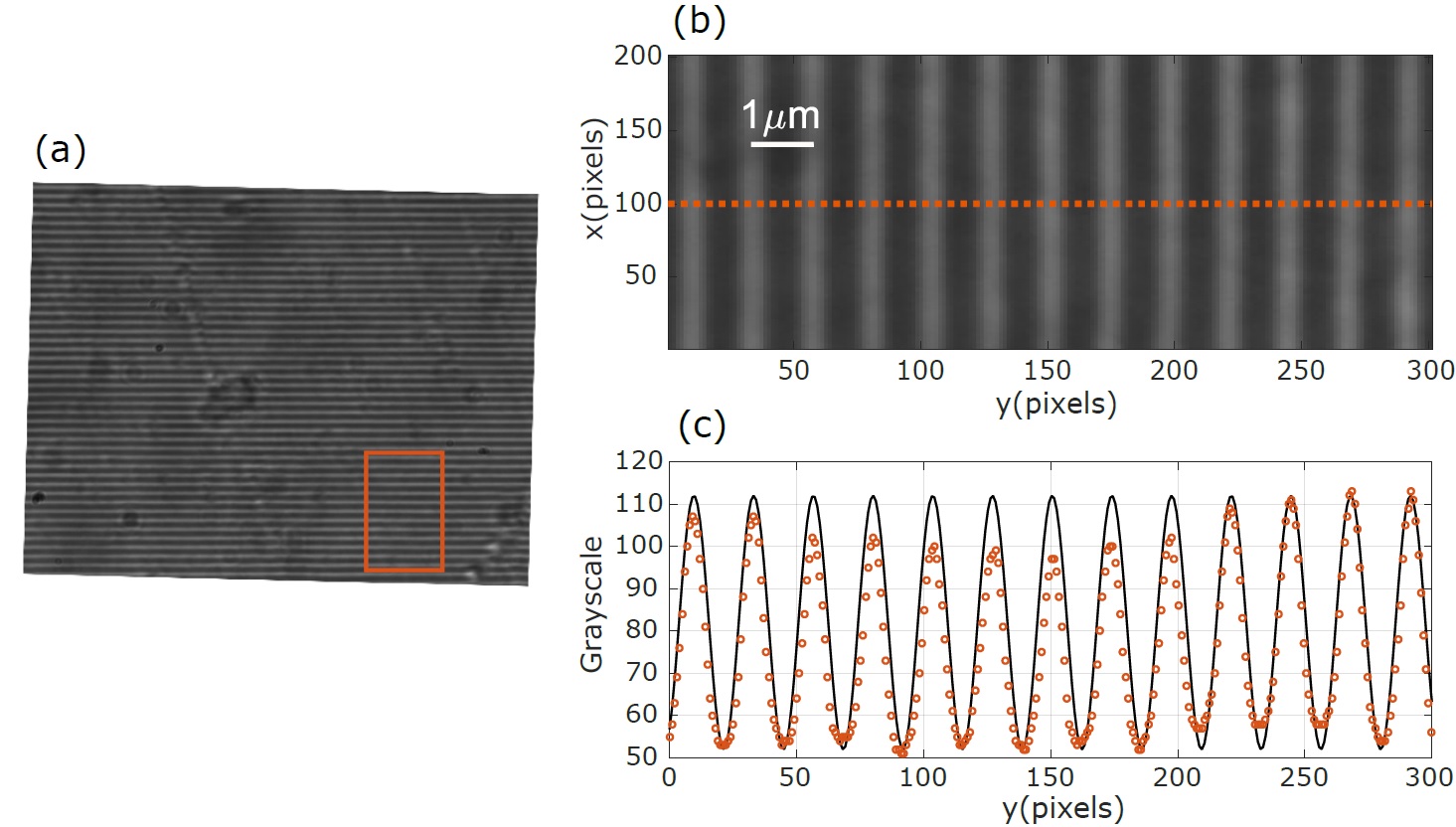}
 \caption{Microscope calibration. $ (a) $ Right image of a $ 1 \mu \text {m} $ periodicity grid made from PDMS material. $ (b) $ rectangular region of the $ 200 \times 300 $ Pixel image. $ (c) $ Grayscale profile taken for line $ x = 100 $. The calibration result is approximately $ L = 23.53 \text{pixels} / \mu \text{m} $.}
 \label{CCD_Calibration}
\end{figure}

When the tracking process is performed, the positions of the particles in each frame is obtained in pixel units. Pixel-to-micrometer calibration was performed by placing a diffraction grating in PDMS (polymeric plastic material) with a frequency of $1 \mu \text{m}$ as can be seen in Figure~\ref{CCD_Calibration} $(a)$. In order to get the pixel-to-micrometer ratio, we chose a rectangular region of the image from $200 \times 300$ Pixel, as shown in Figure \ref{CCD_Calibration} $(b)$. Figure~\ref{CCD_Calibration} $(c)$ shows the grayscale profile taken for the line $x = 100$ (Orange circles). The adjustment is obtained from a sinusoidal function 
 
\begin{equation}
f(x)=A+B\sin\left(\dfrac{2\pi x}{L} - \phi \right) \,,
\label{ajuste}
\end{equation}
where $A$, $B$ are constant, $\phi$ is the phase, and $L$ is the periodicity of the grating. The Fig.~\ref{CCD_Calibration} $(c)$ shown the adjustment obtained from the Eq.\ref{ajuste} (black line). On setting, values obtained for parameters are $A=82.33\pm 2.1\text{pixels}$, $B=30.00 \pm 2.24\text{pixels}$ , $ \phi = 57.29 ^ {0} \pm 0.06 ^ {0} $, and $ L = 23.53 \pm 0.68 \text{pixels} $. Therefore, we can conclude that the calibration factor is $ 0.0425 \mu \text{m} / \text{pixel} $. 

\textbf{Determination of elastic constant $\kappa$}. Since the $x(t)$ path of the time-adjusted particle determined, it is possible to obtain a quantitative measure of the force and optical potential by measuring the elastic constant $\kappa$, also known as stiffness or trap rigidity~\cite{garcia2018high}. There are several calibration methods for measuring the elastic constant, among which the most widely used are: potential analysis from the equipartition theorem, auto--correlation function analysis, and power spectrum analysis~\cite{Jones2015}. On the other hand, a new method (FORMA), introduced by Laura Perez et al.~\cite{garcia2018high}, allows the reconstruction of microscopic optical force using the Maximum Likelihood Method~\cite{kutner2005applied} and assuming that near the equilibrium position the force has a linear behavior with the position. In the following, we describe the FORMA (\textit{Force Reconstruction via Maximum-likelihood-estimator Analysis}) method, that we will use in our work to calculate the elastic constant $\kappa$.   
\\

\textbf{FORMA Method}. The trajectory of a Brownian particle in the presence of a potential $ U(\textbf{r})$ can be modeled in two dimensions by the overdamped Langevin equation~\cite{Jones2015, volpe2013simulation}

\begin{equation}
\dfrac{d}{dt}\textbf{r}(t)=\dfrac{1}{\gamma}\textbf{F}(\textbf{r})+\sqrt{2D}\textbf{W}(t)\,,
\label{Langevin_Equation}
\end{equation}

where $ \textbf{r}(t) = (x(t), y(t)) $ is the position of the particle, $ \textbf{F(r)} $ corresponds to the force field, $ \gamma $ is the particle friction coefficient, which for a spherical particle of radius $ R $ that moves in a viscosity fluid $ \eta $, is determined by Stokes law $ \eta = 6 \pi \gamma R $. $ D = k_{B} T / \gamma $ is the diffusion coefficient, and $ \textbf{W} (t) $ is the noise vector that represents the effect of a continuous series of collisions with atoms of the fluid and is characterized by the following property: $ <\textbf {w} (t)> = 0 $ for each value of $t$~\cite{Jones2015}.
\\

For $N$ particle offset measurements $\Delta \textbf{r}_{n}$ at the position $\textbf{r}_{n}$ during the time interval $\Delta t_{n}$, for $n=1, ..., N$, so the frictional force at $n$--th interval can be written as

\begin{equation}
f_{n}=\gamma\dfrac{\Delta \textbf{r}_{n}}{\Delta t_{n}}=\textbf{J}_{0}\textbf{r}_{n} + \sigma \textbf{w}_{n}\,,
\label{Forca_Atrito}
\end{equation}
where the $ \textbf{F} (\textbf{r}) $ force field was expanded in a Taylor series around the equilibrium position

\begin{equation}
\textbf{F}(\textbf{r})=\textbf{F}_{0}+\textbf{J}_{0}\textbf{r}+O(\textbf{r})\,,
\label{Forca_Atrito}
\end{equation}   
where $ \textbf{F}_{0} = \textbf{F}(0) = 0 $ e $ \textbf{J}_{0} = \textbf{J} (0) $ is the force and the Jacobian in the equilibrium position respectively. $ \sigma = \sqrt {2k_{B} T / \gamma \Delta t_{n}} $, and $ \textbf{w}_{n} $ is a random number array with a value of $ 0 $ and variance $ 1 $. Using the Maximum Likelihood Method ~\cite{kutner2005applied}, the value of $ \textbf{J} _ {0} $ that maximizes the likelihood function in the corresponding linear regression and given by~\cite{garcia2018high}

\begin{equation}
\textbf{J}_{0}^{\ast}=\left[\textbf{r}^{\text{T}} \textbf{r} \right]^{-1} \textbf{r}^{\text{T}}\textbf{f}\,,
\label{Forca_Atrito}
\end{equation}  

where $ \textbf {J}_{0} ^ {\ast} $ is the estimate for $ \textbf{J}_{0} $ per maximum likelihood, and the estimated force field around the equilibrium position is $ \textbf{F}^{\ast} = \textbf{J}_{0}^{\ast} \textbf{r} $. $ \textbf{r} = (\textbf{r}_{n}) $, and $ \textbf{f} = (\textbf{f}_{n}) $ are matrices of $ N \times 2 $. According to Helmholtz's theorem, any force field can be written as a superposition of a conservative and a non-conservative force field. You can get these two components straight from the Jacobian, that is: $ \textbf {J}_{0}^{\ast} = \textbf {J}_{c}^{\ast} + \textbf{J}_{r}^{\ast} $, where the conservative part is given by~\cite{garcia2018high}

\begin{equation}
\textbf{J}_{c}^{\ast}=\frac{1}{2}\left (\textbf{J}_{0}^{\ast}+\textbf{J}_{0}^{\ast \text{T}} \right )=\text{\textbf{R}}\left (\theta^{\ast} \right )\begin{bmatrix}
-k_{1}^{\ast} & 0 \\ 
 0 & -k_{2}^{\ast} 
\end{bmatrix}
\text{\textbf{R}}^{-1}\left (\theta^{\ast} \right ) \,,
\label{Forca_Atrito}
\end{equation}  
where $ \text {\textbf{R}} \left (\theta ^ {\ast} \right) $ is the rotation matrix that diagonalizes $ \textbf{J}_{c} $.In this new coordinate system, where the matrix $ \textbf{J}_{c} $ is represented by a diagonal matrix, the main axes are the eigenvectors corresponding to the main axis of the harmonic potential, and the elastic constants along these axes correspond to to the eigenvalues~\cite{garcia2018high}.
\\

Estimated particle diffusivity along each axis can be calculated from the residual error of the maximum likelihood method:

\begin{equation}
\begin{split}
D_{x}^{*}&=\dfrac{1}{N}\sum_{n=1}^{n}\dfrac{\Delta t_{n}}{2\gamma^2}\left( f_{x,n}-J_{0,11}^{*}x_{n} -J_{0,12}^{*}y_{n} \right)^2\,, \\ 
D_{y}^{*}&=\dfrac{1}{N}\sum_{n=1}^{n}\dfrac{\Delta t_{n}}{2\gamma^2}\left( f_{x,n}-J_{0,21}^{*}x_{n} -J_{0,22}^{*}y_{n} \right)^2. 
\end{split}
\label{difusividade}
\end{equation} 

Comparing with the expected value $ D $, we can have an intrinsic quantitative check of consistency for the quality of the estimate. \\

\h {\em\bf 4. Results and Discussion}  

Initially, it is important to study how the propagation properties of a AiBAs change as it passes through the two $4f$ systems, Fig.~\ref{Setup_Pinza} and was used an Argon laser with wavelength $\lambda = 514.5 \text{nm}$. When the AiBA optically reconstructed by SLM passes through the two $4f $ systems, its propagation properties change according to the Eq.~\eqref{Airy_ABCD}. For this system, the ABCD matrix is given by $M=M_{ij}$ where $A=M_{11}=f_{2}f_{4}/f_{1}f_{3}$, $B=M_{12}=f_{1}f_{3}z/f_{2}f_{4}$, $C=M_{21}=0$ and $D=M_{22}=f_{1}f_{3}/f_{2}f_{4}$.
\\

The characteristic length $w_{0}$ changes according to the relation $(f_{2} f_{4} w_{0}) / (f_{1} f_{3})$, whereas the ballistic trajectory, and the focal point through relationships
\begin{equation}
x=\left( \dfrac{f_{1}f_{3}}{f_{2}f_{4}}\right)^2 \left(\dfrac{z^2}{4k^2 x_0^3} \right) + \left( \dfrac{f_{1}f_{3}}{f_{2}f_{4}}\right)\left(\dfrac{\nu z}{kx_{0}} \right)\,, 
\label{Balisctic_Trajectory_4f}
\end{equation} 
\begin{equation}
z_{f}=-2kw_{0}\left( \dfrac{f_{2}f_{4}}{f_{1}f_{3}}\right)\left(\nu - \sqrt{\nu^2+\dfrac{ s_{0}}{w_{0}}} \right).
\label{Focal_Point}
\end{equation} 

To investigate experimentally how the autofocusing properties of the beams change after propagating through two $4f$ systems, and since the position of the objective and camera on the setup of Fig.~\ref{Setup_Pinza} is fixed for image acquisition on each plane, we generated a sequence of CGHs of AiBAs for different positions along the $z$--axis with $\Delta z=0.02 \text{cm}$. For each position is generated a CGHs, in total $150$ holograms (frames). When having the set of successive images of the field in different positions, these are grouped using the software (\emph{Slideshow player}) of SLM--LETO, facilitating the exhibition of CGHs and the sequence of CGHs in SLM. The CGH video was made at $10$~frames/second. The data acquisition process, in this case, is done keeping the static CCD camera in a given position $z_{0}$ (beginning of the propagation axis) where the image capture (in this case AiBAs) of all the frames with CGHs (AiBAs holograms in each $z$).
\\

Thus, based on this method we experimentally reconstruct each of the generated optical beams (in this case Airy beam overlays) as they pass through the two $ 4f $ systems of Figure \ref{Setup_Pinza}. As well, we could compare with the theoretical simulations of these beams and validate the reliability of the beam reduction process by the $ 4f $ systems in the holographic optical tweezers arrangement we developed for the measurements.
\\

For the optical trapping of the microparticles, let us consider a superposition  of $N = 128$ AiBs symmetrically rotated over the transverse plane accelerating each other in opposite directions. In the input plane ,$z=z_{\text{SLM}}$, the AiBs are characterized by the following parameters: $\lambda=514.5\text{nm}$, $a=0.05$, $w_{0}=50\mu \text{m}$, and $\delta x = \delta y = 200 \mu \text{m} $. \\

\begin{figure}[H]
 \centering
  \includegraphics[scale=0.40]{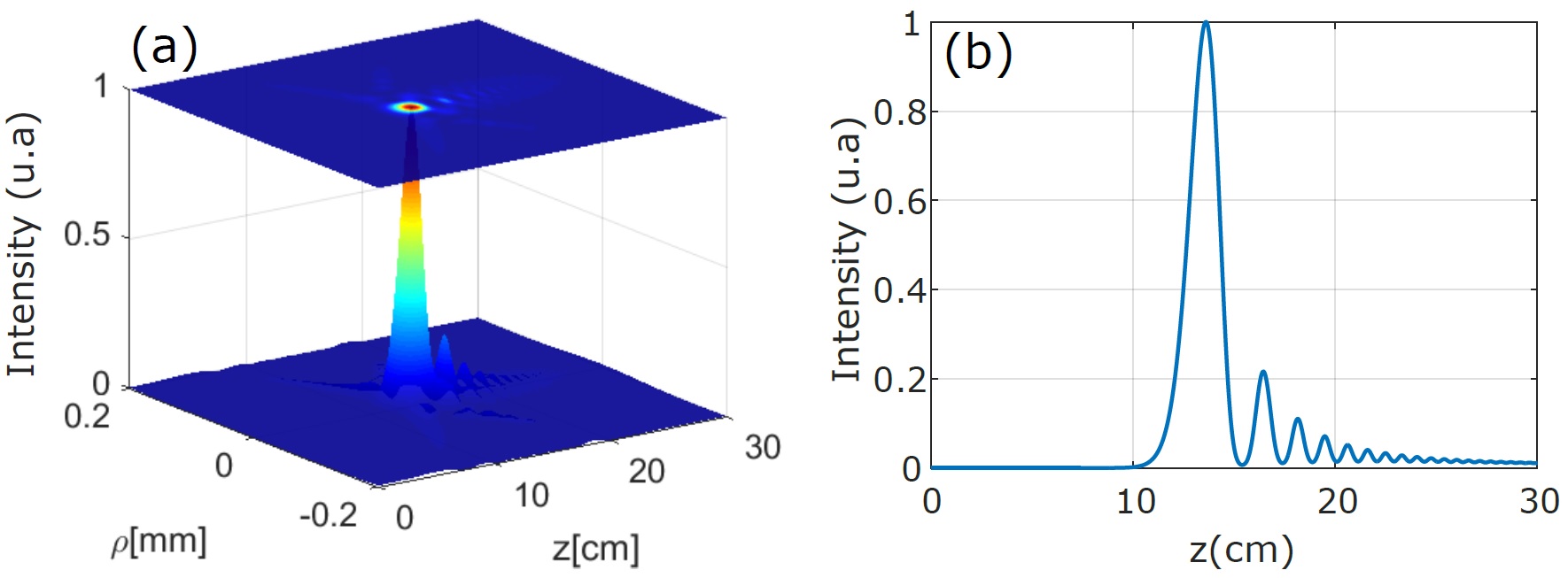}
 \caption{$(a)$ Three-dimensional profile and orthogonal projection of the theoretical normalized intensity generated in the $ z = z_{\text{SLM}} $ plane. $ (b) $ longitudinal intensity pattern (on - axis).}
 \label{Airy_Original_nu=0}
\end{figure}

\textbf{Trap without initial launch angle $ (\nu = 0) $}. Initially, we consider that the beams propagate without an initial launch angle, $\nu=0$. The three-dimensional profile and orthogonal projection of the theoretical normalized intensity generated at SLM output can be seen in Fig.~\ref{Airy_Original_nu=0}~$(a)$. The corresponding longitudinal intensity pattern along the on--axis is shown in Fig.~\ref{Airy_Original_nu=0}~$(b)$. We can see that the beams converge on the $z_{f}=13.6\text{cm}$ plane. \\

\begin{figure}[H]
 \centering
  \includegraphics[scale=0.65]{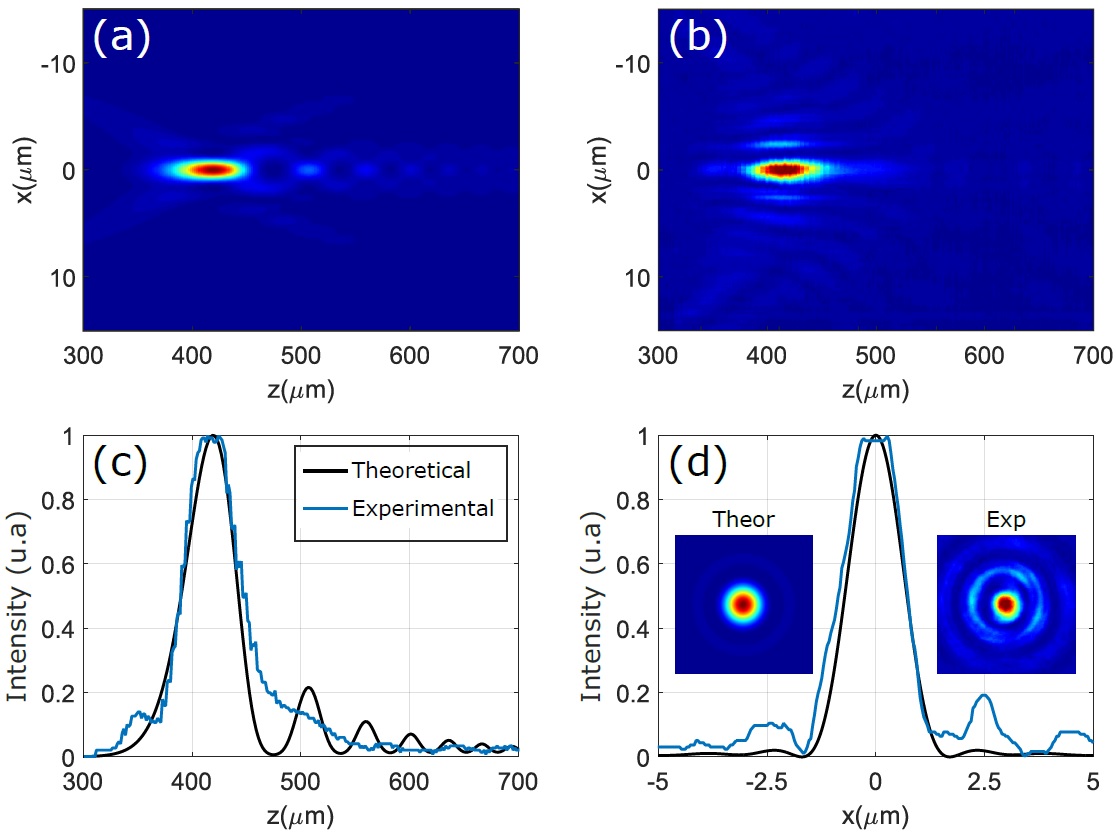}
 \caption{$ (a) $ Orthogonal projection of theoretical normalized intensity for an arrangement of N $ = 128 $ Airy bundles and initial launch angle $ (\nu = 0) $ when passing through two $ 4f $ systems. $ (b) $ Experimental result obtained through the arrangement \ref{Setup_Pinza}. $ (c) $ Comparison of longitudinal intensity pattern between theoretical prediction (black line) and experimental result (blue line). $ (d) $ Cross-sectional intensity profile of the plane $ z_{f} = 418 \mu m $.}
 \label{Airy_Transversal_nu=0}
\end{figure}

Figure~\ref{Airy_Transversal_nu=0}~$(a)$ shows the orthogonal projection of the theoretical intensity as the beam passes through the $4f$ systems of the experimental arrangement of Figure~\ref{Setup_Pinza}. The corresponding experimental result is shown in Figure~\ref{Airy_Transversal_nu=0}~$(b)$. After passing through $4f$ systems the focal point will be $z_{f}=418\mu\text{m}$. In Figure~\ref{Airy_Transversal_nu=0}~$(c)$ we compare the intensity profile along the propagation axis ( $z$--axis), theoretical (black line) and experimental (blue line) axis. The normalized intensity cross--section at the focal point $z_{f}$ is shown in Figure~\ref{Airy_Transversal_nu=0}~$(d)$, where the size of spot radius is $\Delta \rho=2.2 \pm 0.1 \mu\text{m}$. It is important to note that the size of the spot obtained experimentally is very close to the theoretical result.
\\

\begin{figure}[H]
 \centering
  \includegraphics[scale=0.65]{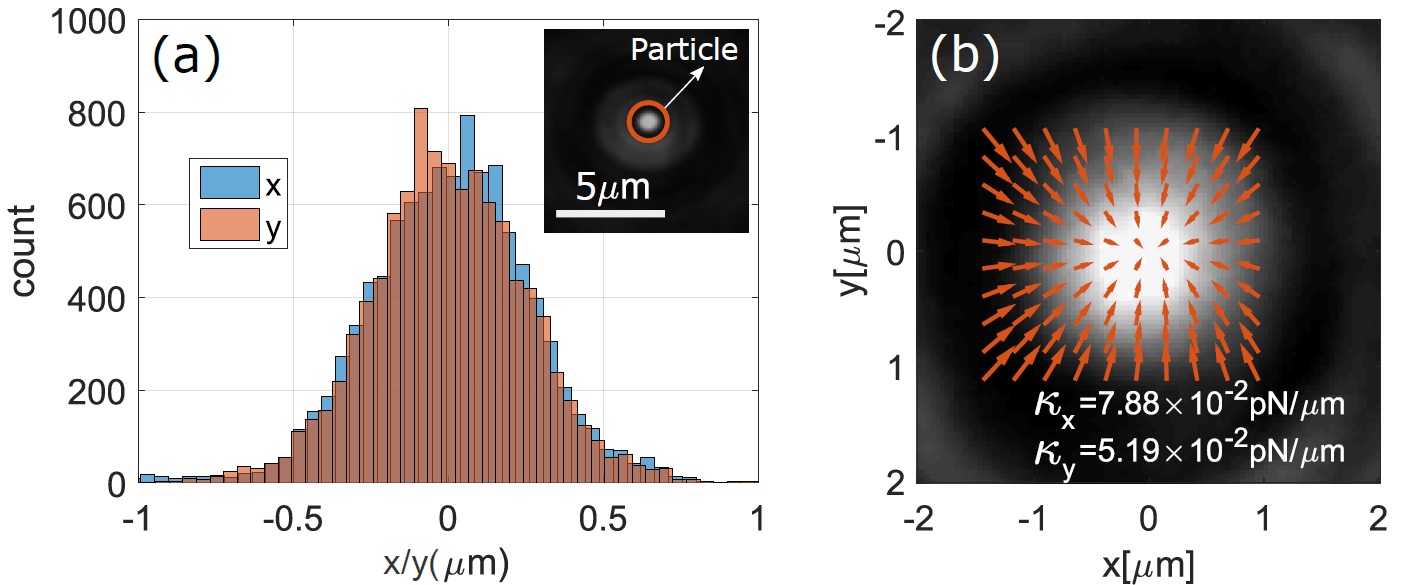}
 \caption{$(a)$ Position histograms for the $x$--axis (blue color), and --$y$ (red color) for an optically trapped particle (orange circle) at $z_{c}$ (Visualization 1). $(b)$ Force field reconstruction from the FORMA method, where $\kappa_x$ and $\kappa_y$ are elastic constant in x and y-axis, respectively.}
\label{Histogram_nu=0}
\end{figure}

Implementing this beam in the holographic optical tweezers, we have in Figure~\ref{Histogram_nu=0}~$(a)$ the distribution of the microparticle position (orange circle) transversely trapped by the gradient force at in the focal plane $z_{f}$ which corresponds to a stable equilibrium point. The power in the sample plane was $ 2.5 \pm 0.3 \text{mW}$. To perform the tracking, the trapped particle was recorded for $10000$ frames where the time between each frame was $5.7\text{ms}$, and the FORMA method was used to determine the force distribution. We can observe how the particle moves around the equilibrium position $ x_{\text{eq}} = 0 $, thus verifying that the distribution of the particle position in the trap. The force field distribution in the $ z_{f} $ plane can be seen in Fig.~\ref{Histogram_nu=0}~$(b)$. The direction and magnitude of the arrows (orange color) correspond to the direction and magnitude of the transverse force, where $\kappa_{x}=7.88\times 10^{-2}\text{pN/}\mu\text{m} $ and $\kappa_{y}=5.19\times 10^{-2} \text{pN/}\mu\text{m}$. The intensity of the experimental cross--section of the beam in the trapping background plane is shown of the Fig.~\ref{Histogram_nu=0}~$(b)$.
\\

\textbf{Trap with initial launch angle $ (\nu \neq 0) $}. 

\begin{figure}[H]
 \centering
  \includegraphics[scale=0.60]{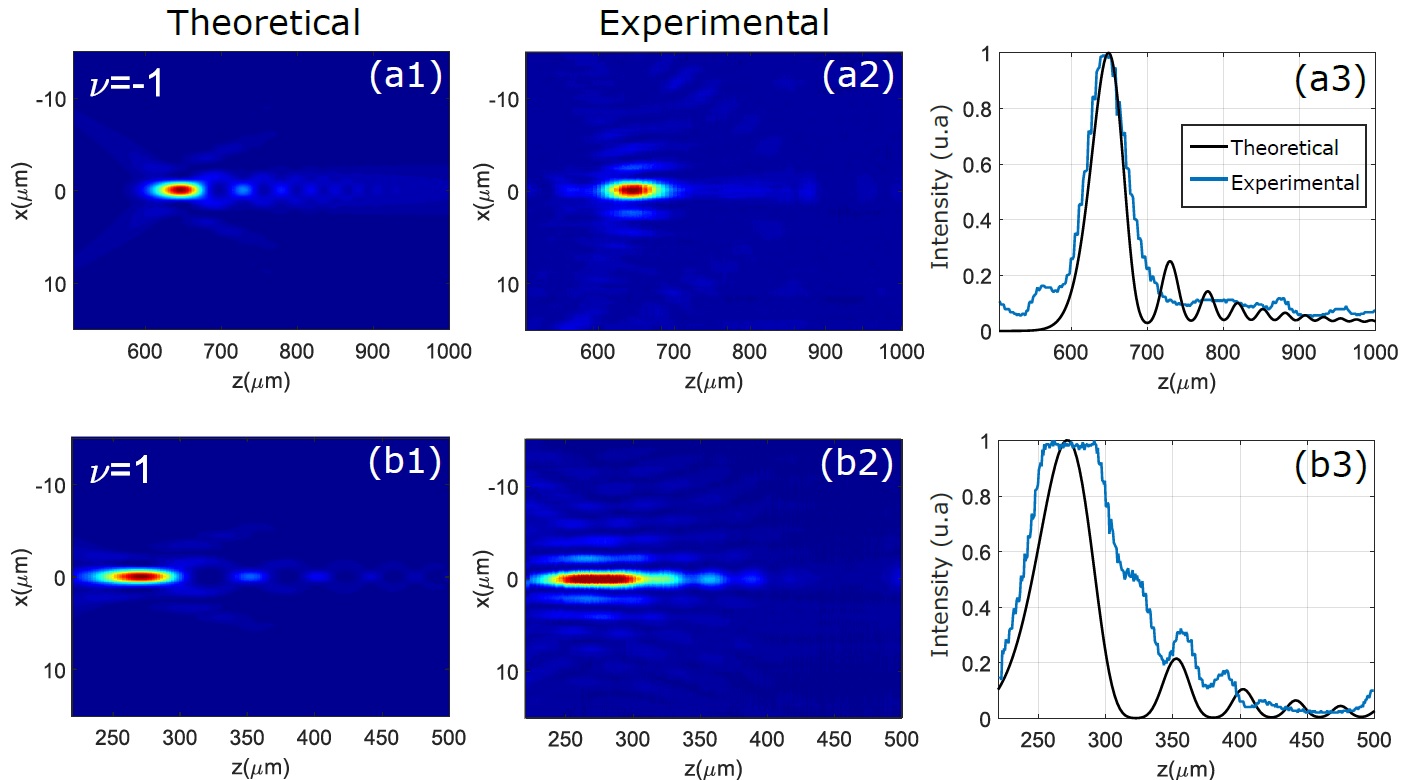}
 \caption{$(a1)$ e $(b1)$ Orthogonal projection of theoretical normalized intensity with initial pitch angle $(\nu=-1)$ and $(\nu=1)$ when passing two systems $4f$. $(a2)$ and $(b2)$ Experimental result. $(a3)$ and $(b3)$ Comparison of the normalized longitudinal intensity profile between the theoretical (black line) and the experimental result (blue line) for $\nu=-1$ and $\nu=1$, respectively.}
 \label{Airy_Propagation_different_nu}
\end{figure}

\begin{figure}[H]
 \centering
  \includegraphics[scale=0.60]{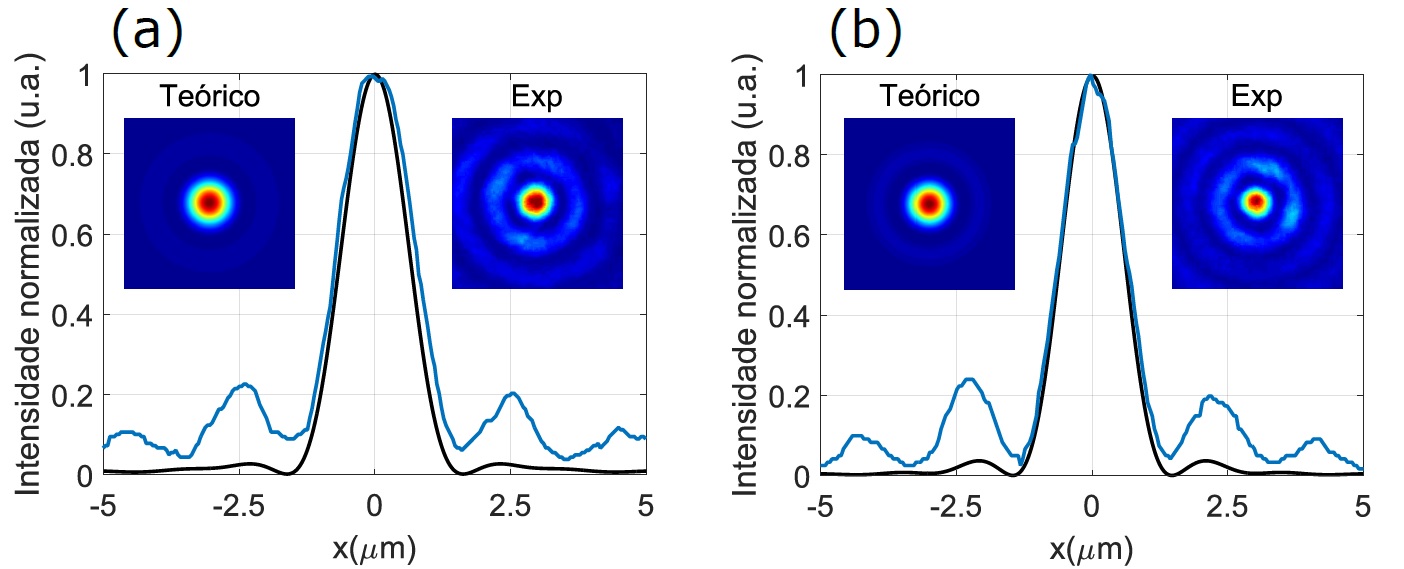}
 \caption{Normalized intensity cross--section over the focal plane. $(a)$ $\nu=-1$ in $z_{f}=649.3\mu\text{m}$. $(b)$ $\nu=1$ in $z_{f}=270.5\mu\text{m}$.}
 \label{Airy_Transversal_different_nu}
\end{figure}

Let us now consider that each AiB in the array propagate with the same initial launch angle $(\nu \neq 0)$. Fig.~\ref{Airy_Propagation_different_nu}~$(a1)$ and $(b1)$ shows the orthogonal projection of the theoretical normalized intensity when the AiBAs passes through the $4f$ systems for $\nu=-1$ and $\nu=1$, respectively. The corresponding experimental result obtained from the arrangement of Fig.~\ref{Setup_Pinza} can be seen in Fig.~\ref{Airy_Propagation_different_nu}~$(a2)$ and $(b2)$. In Fig.~\ref{Airy_Propagation_different_nu}~$(a3)$ and $(b3)$, we compare the normalized intensity pattern along the propagation ($z$--axis), theoretical (black line), and experimental (blue line). This set of beams converge at the points $z_{f}=649.3 \mu \text{m}$ and $ z_{f}=270.5 \mu \text{m} $ to $\nu=-1$ and $\nu=1 $, respectively. The normalized cross--section intensity profile in the focal planes is shown in Fig.~\ref{Airy_Transversal_different_nu}.  In both case, the $spot$ radius is very close to the theoretical result.

\begin{figure}[H]
 \centering
  \includegraphics[scale=0.50]{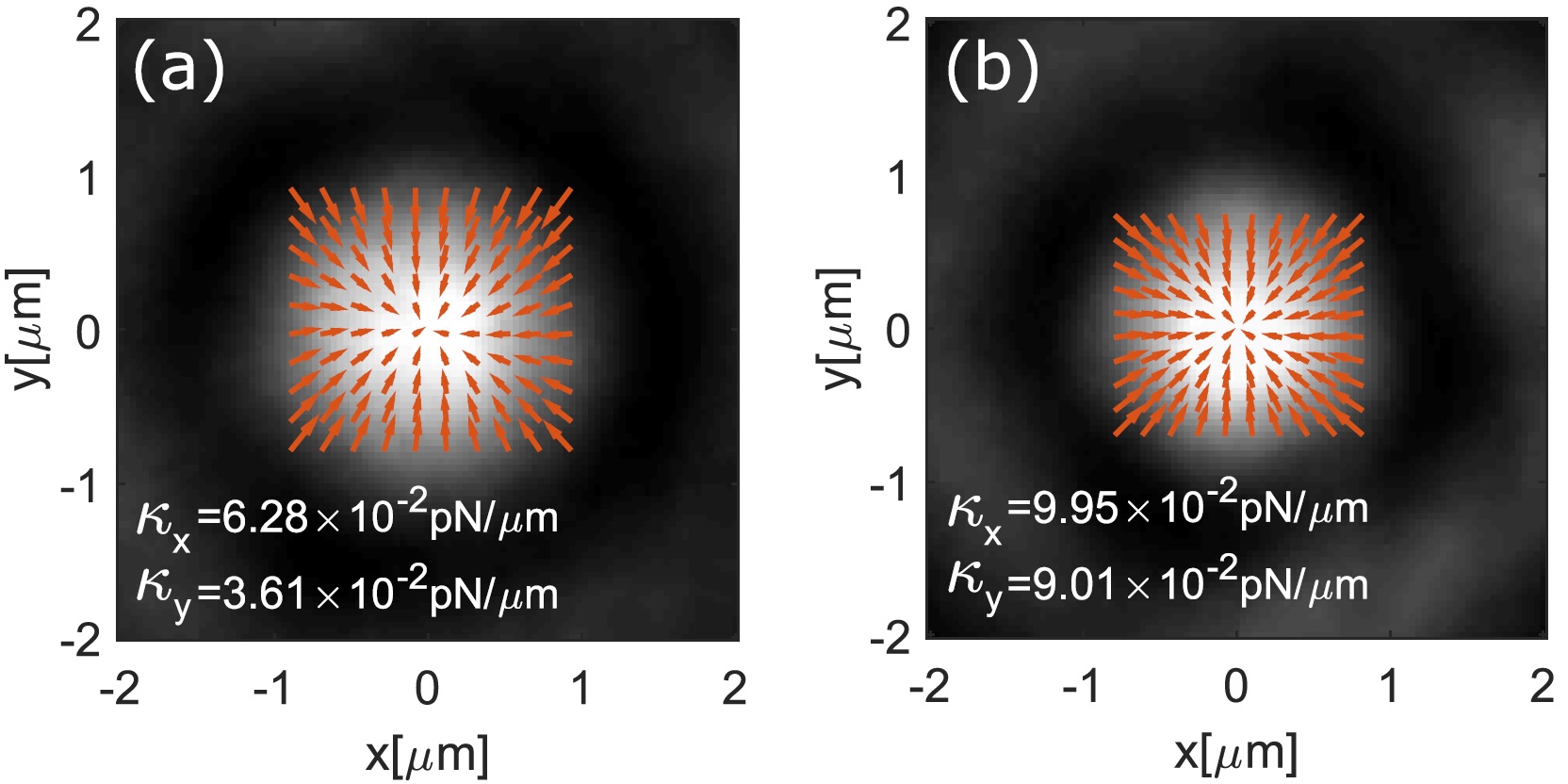}
 \caption{Force field reconstruction using the Forma method. Direction and magnitude (orange arrows) corresponds
to the direction and magnitude of the transverse force at the plane $z_{f}$. $(a)$ $\nu=-1$ in $z_{f}=649.3 \mu \text{m}$ (Visualization 2). $(b)$ $\nu=1$ in $z_{f}=270.5 \mu \text{m}$ (Visualization 3). }
 \label{Airy_Force_Distribution_Different_nu}
\end{figure}

\begin{figure}[H]
 \centering
  \includegraphics[scale=0.50]{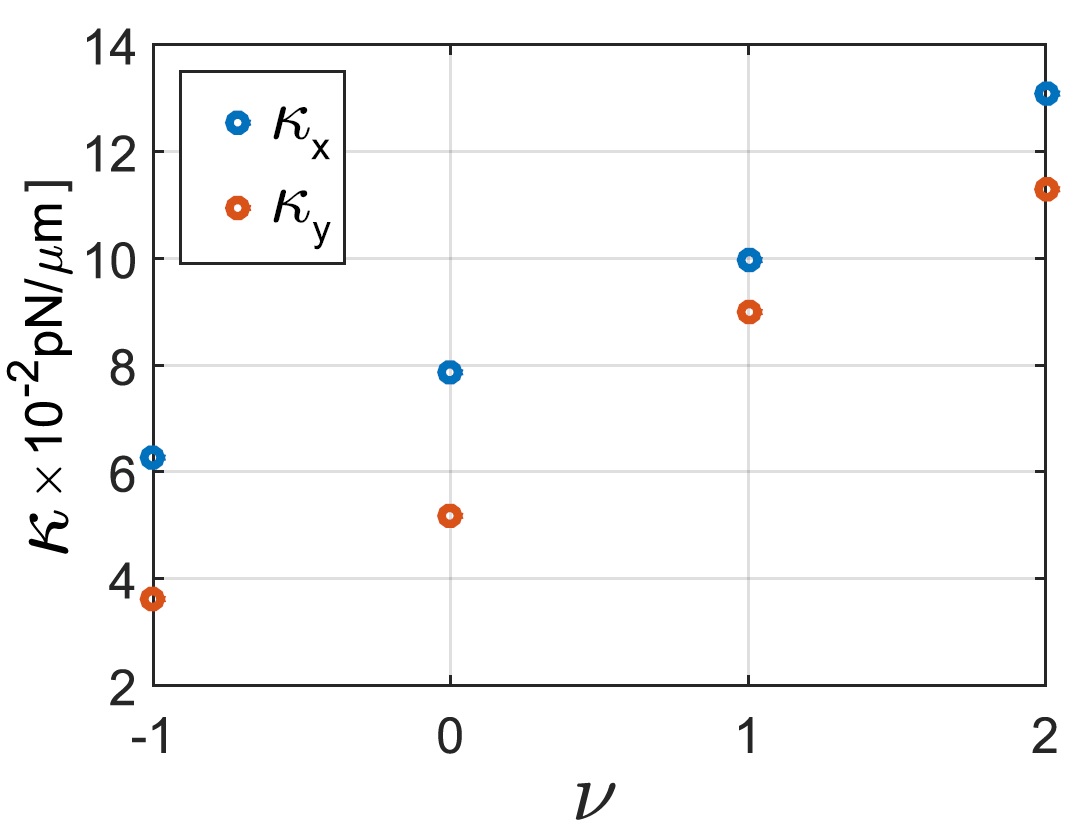}
 \caption{Trap stiffness $\kappa_{x}$ and $\kappa_{y}$ as a function of the initial launch angle $\nu$.   }
 \label{Stiffnes_Different_nu}
\end{figure}

From the experimental generation of AiBA with different initial launch angles it was possible to trap  micro--particles with different values of $\nu$; $\nu=-1$ and $\nu=1$ in the $z_{f}=649.3\mu\text{m}$ and $z_{f}=270.5\mu\text{m}$ plans, respectively. Fig.~\ref{Airy_Transversal_different_nu} shows the force field distribution for different values of $\nu$. The direction and magnitude of the arrows (orange) correspond to the direction and magnitude of the transverse force. The intensity  of the experimental cross--section in the capture plane is shown in the background of each figure. We can observe that for $\nu=1$ the particle movement around the equilibrium position is much smaller than for $\nu=0$ and $\nu=1$, thus having a greater stiffness in the trap for positive values of $\nu$. Fig.~\ref{Stiffnes_Different_nu} shows the elastic constant ($\kappa_{x}$ and $\kappa_{y}$) as a function of initial launch angle $\nu$, where we can see how it can be increased by choosing positive values of $\nu$. 
\\

\textbf{Trapping and guiding via dynamic holography}. 

We can study how microparticles can be trapped and guiding on transverse planes. In this case, we use a dynamic sequence of computer-generated holograms to create a dynamic scenes~\cite{Vieira2015,Rached2018} and experimentally reproduce the movement of the beam on the transversal plane.
\\

\begin{figure}[H]
 \centering
  \includegraphics[scale=0.60]{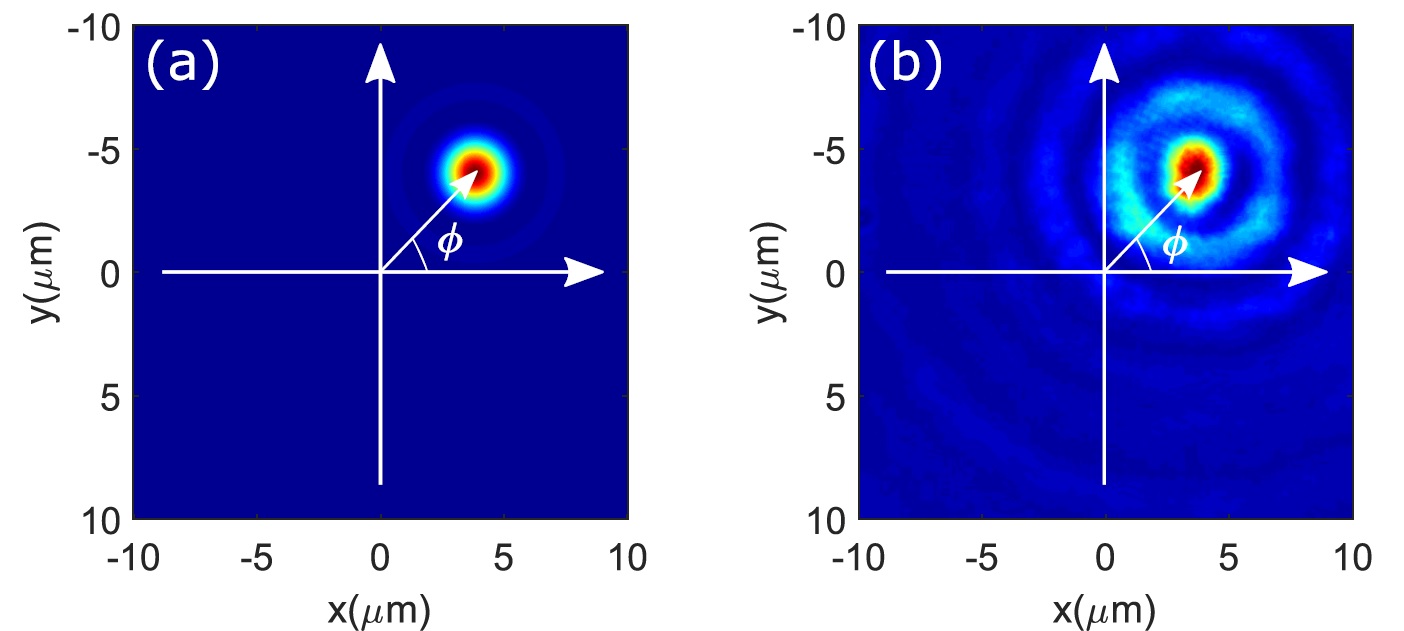}
 \caption{Cross–sectional intensity profile in the focal plane $z_{f}=418\mu m$ for $\phi=\pi/4$. $(a)$ Theoretical and $(b)$ Experimental.}
 \label{Rotation_Trapping}
\end{figure}

We consider that the trap is formed by an AiBA of N$=128$ Airy beams-$2D$ symmetrically rotated on the transverse plane. In the initial plane $z=z_{\text{SLM}}$ the Airy beams are characterized by the following parameters: $\lambda=514 ~\text{nm}$, $ a=0.05$, $\nu=0$, $w_{0}=50 \mu \text{m} $, and $\delta x=\delta y =200\mu \text{m}$. We now consider the origins of coordinates displaced spatially on the transverse plane to the direction of propagation, that is,
\begin{equation}
x^{'}=x+r\cos \phi, \quad y^{'}=y+r\sin \phi\,,
\end{equation} 
where $r=100\mu \text{m}$ and $\phi$ can take values between $0$ and $2\pi$.  When this beam passes through the two $4f$ systems in Fig.~\ref{Setup_Pinza}, the propagation properties change according to the equation~\eqref{Airy_ABCD}. Fig.~\ref{Rotation_Trapping} shows the theoretical and experimental cross--section of the intensity profile in the focal plane $z_ {f}=418\mu \text{m}$ for $\phi = \pi/4$.
In order to observe how microparticles can be trapped in the focal plane $z_{f}$ and guided over the transversal plane $x$-$y$, we generate a sequence of holograms of the beam in different positions in order to create a dynamic scene. Each hologram is generated by rotating the beam at an angle $\phi$ and keeping $r=100 \mu\text{m}$ constant. In this case, we take $\phi=N\pi/50 $ where $N=100$ is the number of holograms. In dynamic SLM (LETO, Holoeye Photonics), the frames are grouped to generate the dynamic scene on the transversal plane. The complete rotation was made in $50$ seconds with a velocity of $2$ holograms/second. The video was recorded at $32$Hz so that the time between each frame is $1/32 \text{s} \approx 31.3 \text {ms} $.
\\

\begin{figure}[H]
 \centering
  \includegraphics[scale=0.60]{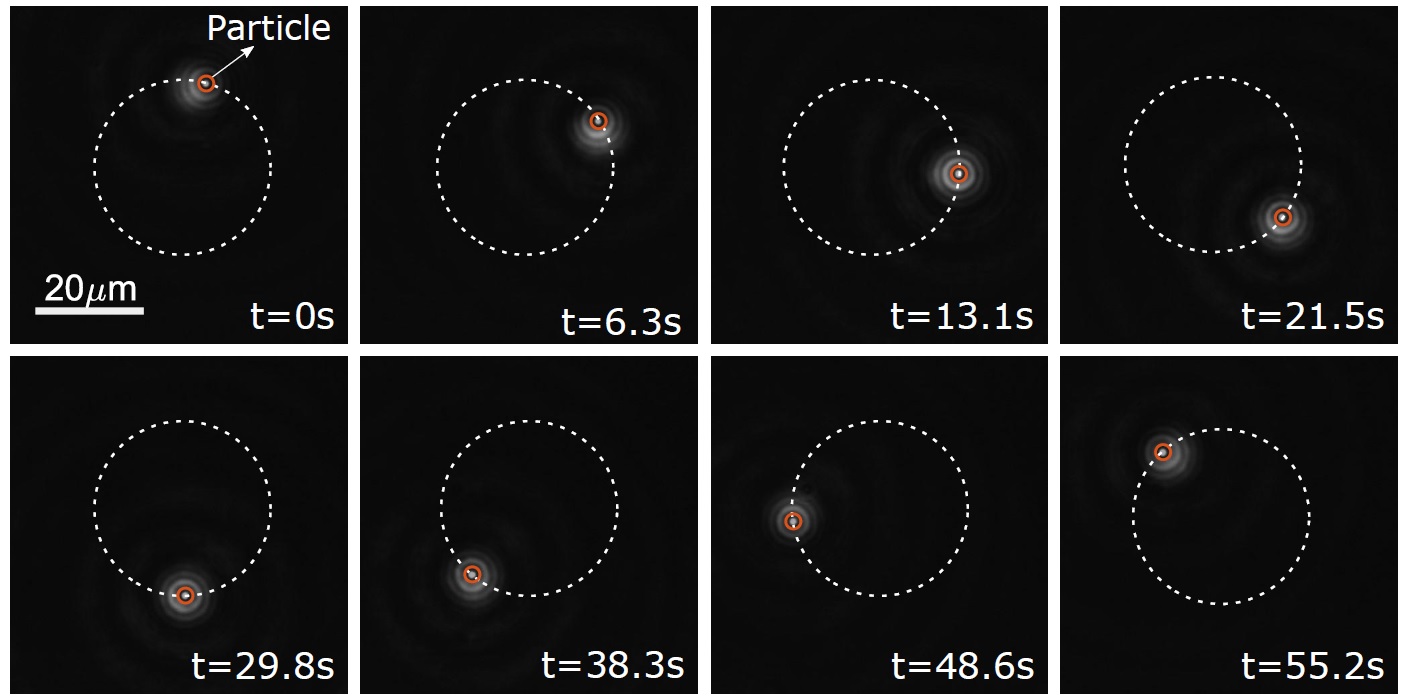}
 \caption{Trapping and guidance in the transverse plane $z_{f}$ of a single microparticle (orange circle) when the AiBA moves around a circumference at a constant angular velocity (Visualization 4).  The power in the capture plane was $1.8 \pm 0.5\text{mW}$.}
 \label{Dynamic_Trapping}
\end{figure}

Figure~\ref{Dynamic_Trapping} shows how the microparticle is trapped at the same time that it is guided on the transverse plane. In this case, the particle is stably trapped and remains in the trap when guided around a circumference of radius $5.5\mu \text{m}$ while the trap moves at a constant angular velocity. These results suggest that the holograms generated by computer allow to create dynamic orientable traps with a high pressure for the trapping and guiding of microparticles.

\h {\em\bf 5. Conclusions}  

In summary, we developed a holographic optical tweezers system for the generation of non--diffracting beams and with it we investigate the distribution of optical forces acting on microparticles of an Airy beams array. The experimental optical forces of AiBA with different initial launch angles $\nu$ exerted on microparticles are investigated. Results indicate a significant improvement in trap stability by increasing the initial angle $\nu$. This increase allows the focal plane of the AiBAs to be reduced, increasing the intensity of the field and greater gradient forces at this point. On the other hand, the control of the $\nu$ parameter allows increased AiBA convergence and larger gradient  forces, similar to using a higher NA objective in conventional optical tweezers. One of the advantages of optical trapping with AiBAs with different initial launch angles over conventional optical tweezers is the self--focusing properties of AiB.  This allows us to have optically a transverse and longitudinal control of the trap, generating a dynamical optical tweezers for guiding and trapping in 3--D. Therefore, our results with a superposition of AiBs with self--focusing properties brings  interesting possibilities for trapping and guiding of microparticles in a controllable way that can be applied in optical, biological and atmospheric sciences.  \\

\h {\em Acknowledgments.} The authors would like to thank Tarcio A. Vieira and Michel Zamboni-Rached for fruitful technical discussions. This work was supported by Federal University of ABC (UFABC); Coordenaç\~ao de Aperfeiçoamento de Pessoal de N\'ivel Superior-CAPES; Fundaç\~ao de Amparo a Pesquisa do Estado de S\~ao Paulo-FAPESP (16/19131-6); Conselho Nacional de Desenvolvimento Cient\'ifico e Tecnol\'ogico-CNPQ (302070/2017-6).

\end{document}